\documentclass[
]{ceurart}

\usepackage{listings}
\usepackage{subcaption}
\begin{document}

\copyrightyear{2022}
\copyrightclause{Copyright for this paper by its authors.
  Use permitted under Creative Commons License Attribution 4.0
  International (CC BY 4.0).}

\conference{AIQxQIA 2026: International Workshop on AI for Quantum and Quantum for AI | co-located with AIxIA 2026, Perugia, Italy}

\title{Quantum Annealing for Gamma-ray Source Detection and Localization with AGILE Data}


\author[1]{Alessandro Rizzo}[%
  orcid=0009-0003-4341-2988,
  email=alessandro.rizzo@inaf.it,
]
\cormark[1]

\author[2]{Nicolò Parmiggiani}[%
  orcid=0000-0002-4535-5329,
]

\author[2]{Andrea Bulgarelli}[%
  orcid=0000-0001-6347-0649,
]

\author[2]{Gabriele Panebianco}[%
  orcid=0000-0002-3410-8613,
]

\author[3]{Carlo Burigana}[%
  orcid= 0000-0002-3005-5796 ,
]

\author[4]{Vincenzo Fabrizio Cardone}[%
  orcid=0000-0002-2079-7438,
]

\author[1]{Farida Farsian}[%
  orcid=0009-0004-5252-1737,
]

\author[2]{Massimo Meneghetti}[%
  orcid=0000-0003-1225-7084 ,
]

\author[5]{Giuseppe Murante}[%
  orcid= 0000-0002-5155-130X ,
]

\author[4]{Giuseppe Sarracino}[%
  orcid=0000-0001-9512-1961,
]

\author[4]{Roberto Scaramella}[%
  orcid=0000-0003-2229-193X,
]

\author[1]{Francesco Schillirò}[%
  orcid=0000-0001-5106-2277,
]

\author[4]{Vincenzo Testa}[%
  orcid=0000-0003-1033-1340,
]

\author[3]{Tiziana Trombetti}[%
  orcid=0000-0001-5166-2467,
]

\author[1]{Fabio Roberto Vitello}[%
  orcid=0000-0003-2355-3797,
]

\address[1]{INAF -- Osservatorio Astrofisico di Catania, Via Santa Sofia 78, 95123, Catania, Italy}
\address[2]{INAF -- Osservatorio di Astrofisica e Scienza dello Spazio, Via Piero Gobetti 93/3, Bologna, 40129, Italy}
\address[3]{INAF -- Istituto di Radioastronomia,  Via Piero Gobetti 93/3, Bologna, 40129, Italy}
\address[4]{INAF -- Osservatorio Astronomico di Roma, Via Frascati 33, Monteporzio Catone, 00078, Italy}
\address[5]{INAF -- Osservatorio Astronomico di Trieste, Via Tiepolo 11, 34131, Trieste, Italy}

\cortext[1]{Corresponding author.}


\begin{abstract}
The detection and localization of gamma-ray sources via maximum-likelihood analysis is a complex parameter optimization problem. We address this challenge using quantum annealing, a computational approach that solves optimization tasks by encoding them into the ground state of an Ising Hamiltonian. Specifically, we apply this method to observations from AGILE's Gamma-Ray Imaging Detector (GRID) in the 30 MeV-50 GeV band. We reformulated the standard analysis as a Quadratic Unconstrained Binary Optimization (QUBO) problem. We tested this method on a set of AGILE observations generated with AGILEpy, solving the QUBO with simulated quantum annealing. The method reproduces the classical flux and position estimates within their statistical uncertainty and identifies the correct grid node among the candidate positions in 97\% of the maps. These promising results demonstrate that quantum annealing could be used in this optimization problem, laying the ground for its deployment on actual quantum annealers.
\end{abstract}


\maketitle

\section{Introduction}
Gamma-ray sources are associated with the most energetic processes in the universe \citep{2011hea..book.....L}, and their detection and characterization are the primary scientific goal of space-borne gamma-ray telescopes such as AGILE \citep {Tavani_2009} and EGRET \citep{thompson1993}. They detect gamma-ray photons and reconstruct their arrival direction and energy, producing count maps from which sources must be extracted by fitting a physical model. The standard approach for doing so is based on a likelihood function introduced for EGRET \citep{1996ApJ...461..396M}. Maximizing the Poisson likelihood of the observed photons yields the parameters of interest: the fluxes of the sources, the normalizations of the two background components, the source position, and the statistical significance of the detection, making this an optimization problem over the model parameters. 

This kind of task is a natural target for quantum computing, an emerging field of computer science and engineering that leverages the principles of quantum mechanics to solve problems beyond the capabilities of even the most powerful classical computers \citep{NC01,RP14}. In recent years, there have been constant advances in this area, driven by developments in quantum hardware \citep{Roffe_2019, bertels2024quantumcomputingnew, quantum6040039, Gill_2025}. However, we are still in the so-called Noisy Intermediate-Scale Quantum (NISQ) era \citep{Preskill_2018}, and the main goal of this discipline is to achieve quantum utility. Nonetheless, several applications have already been explored \citep{bulgarelli2026qcinaf}. In particular, the use of quantum computing to astrophysics and cosmology is an active and growing line of research \citep{Rizzo2024_ADASS, farsian2025qft, sarracino2025qmcmc, sarracino2026qga, farsian2026qgnn, rizzo2026comparing}.
In this work, we explore whether the gamma-ray likelihood optimization described above can be addressed with quantum annealers \citep{annealing, farhi2000quantumcomputationadiabaticevolution}, such as the superconducting devices developed by D-Wave Systems \citep{d-wave}. The method is validated using simulated AGILE observations of gamma-ray sources of different significance, solved with simulated quantum annealing (SQA), and compared against the classical maximum-likelihood pipeline.

We emphasize that this work is a methodological proof of concept, its purpose is not to compete with the classical pipeline on computational cost yet, but to demonstrate that a widely used classical algorithm can be encoded in quantum form without loss of scientific accuracy.

This paper is organized as follows: Section \ref{sec:classical} reviews the classical likelihood analysis and describes the simulated dataset, Section \ref{sec:quantum} presents the QUBO formulation, Section \ref{sec:res} reports the results, finally Section \ref{sec:conc} presents the conclusions and directions for future work.

\section{Classical Analysis and Simulated Dataset}
\label{sec:classical}
The classical analysis that will be used as a benchmark is based on a likelihood function \citep{1996ApJ...461..396M}, that takes the following form:
\begin{equation}
    \ln L = \sum_{ij} n_{ij} \ln(\theta_{ij}) - \sum_{ij} \theta_{ij},
    \label{eq:placeholder_label}
\end{equation}
where $n_{ij}$ are the observed counts and $\theta_{ij}$ the predicted counts in pixel $(i,j)$.

The predicted number of counts is a linear combination of the emission components: 
\begin{equation}
    \theta_{ij} = g_m G_{ij} + g_b \times 10^{-5} \times E_{ij} + c_a \, \mathrm{PSF}(\alpha, \delta, i, j), \label{eq:placeholder_label}
\end{equation}
where $G_{ij}$ is the galactic diffuse emission model convolved with the response, $E_{ij}$ is the exposure map, and $PSF(\alpha, \delta, i, j)$ is the fraction of the PSF (Point Spread Function) centered at $(\alpha, \delta)$ falling in pixel $(i,j)$. The normalization parameters $g_m$ and $g_b$ scale the galactic and isotropic diffuse emission models, respectively, the parameter $c_a$ is the number of counts of the active source to be determined.

Source significance is assessed through the likelihood-ratio test. The null hypothesis ($H_0)$ assumes no point source at the position under consideration and has two free parameters $(g_m, g_b)$, the alternative hypothesis ($H_1$) adds the source counts $c_a$. The test statistic is then defined as $TS \equiv -2(lnL_0 - lnL_1)$ \citep{wilks1938}. 
It is important to note that in this work we consider the simplified case of a single active source observed at high Galactic latitude ($b = 30^{\circ}$), where the Galactic diffuse component $G_{ij}$ is negligible, so that the corresponding term can be dropped. Therefore, the whole model reduces to two free parameters.

When the source position is unknown, the two coordinates $(\alpha, \delta)$ are treated as additional model parameters. They are estimated by scanning the likelihood over a regular grid of candidate positions, at each node the parameters $g_b$ and $c_a$ are fitted and the corresponding TS is evaluated to find the best-fit position. \\

\subsection{Dataset simulation}
AGILE is a mission of the Italian Space Agency launched in 2007 \citep{Tavani_2009}. Its Gamma-Ray Imaging Detector (GRID) observes in the 30 MeV - 50 GeV band with a PSF of about $10^{\circ}$ above 50 MeV. The dataset is generated with the AGILE Science Tools and the AGILEpy analysis framework \citep{agilepy}. The instrument response, exposure, and Galactic diffuse emission maps are derived from a real five-day AGILE pointing, processed over a $50^\circ$ field of view with a $0.5^\circ$ pixel size in the 100 MeV - 10 GeV energy band. The field is centered at high Galactic latitude (b = 30°), so that, as noted above, only the isotropic background and a single point source contribute.

We generated 500 simulated maps, drawing the true parameters at random for each: the source flux log-uniformly in $[5 \times 10^{-5}, 3 \times 10^{-4}]  \text{ ph cm}^{-2} \text{s}^{-1}$, and the isotropic background $g_b$ uniformly in [5,15]. It is important to note that this is a dataset with high source fluxes. This choice reflects the exploratory nature of this work, whose goal is to understand whether a quantum annealing approach to the gamma-ray likelihood problem is viable at all.

\begin{figure}[!h]
    \centering
    \includegraphics[width=0.6\linewidth]{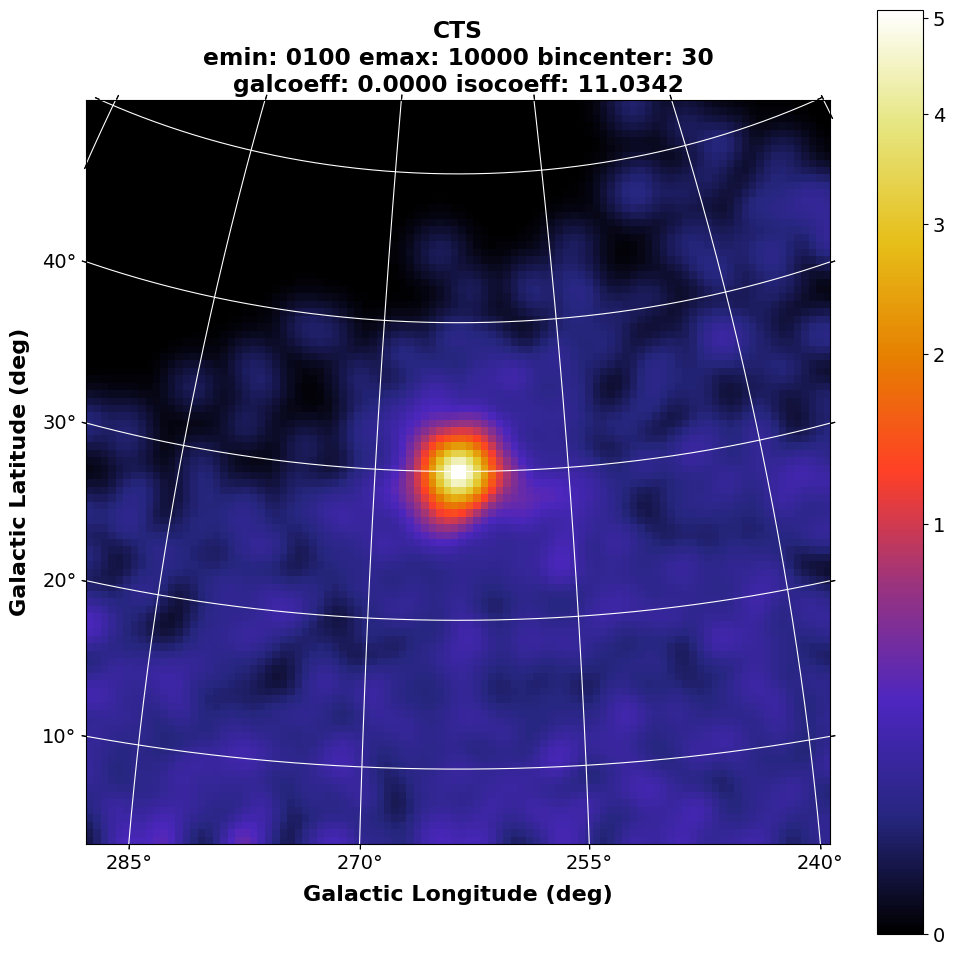}
    \caption{A simulated AGILE count map centered at high Galactic latitude.}
   \label{fig:sim-maps}
\end{figure}

The unit-normalization templates required by both the classical fit and the QUBO are produced by simulating each component in isolation and averaging $N_{avg} = 1000$ independent Poisson realizations. Each template represents the spatial distribution of the expected counts for a single unit of the corresponding parameter. The dictionary of candidate source templates $S_{ij}^{(m)}$ for localization is built in the same way over a regular grid of $M$ = 81 candidate positions around the source location.\\
An example of a simulated count map is shown in Figure \ref{fig:sim-maps}.

\section{QUBO Formulation}
\label{sec:quantum}
Quantum annealers minimize objective functions of the Quadratic Unconstrained Binary Optimization (QUBO) form, which is equivalent to finding the ground state of an Ising Hamiltonian. However, the Poisson log-likelihood is not polynomial, and its parameters are continuous rather than binary.

The first obstacle is resolved by the Gaussian approximation, reducing the log-likelihood to the least-squares cost:
\begin{equation}
    \label{eq:placeholder_label}
    \chi^{2} = \lVert Rx - d \rVert^{2} = x^{T} W x - 2 V^{T} x + d \cdot d,
\end{equation}
where $x = (g_b, c_a)^T$ is the parameter vector, $d$ is the flattened counts map, and $R$ is the matrix whose columns are the component templates ($B_{ij}$ for the isotropic background component, $S_{ij}$ for the source), so that the model prediction is $\theta = Rx$. The matrix $W = R^TR$ encodes the spatial overlap between templates, while the vector $V = R^Td$ measures the alignment of the data with each template.

A fixed-point binary expansion solves the second obstacle. Each continuous parameter $x_i \in [0, x_{max}]$ is written as $x_i = \beta\sum^{n-1}_{k=0}2^k q_k$ with $q_k \in \{0,1\}$ and resolution $\beta = x_{max}/(2^n - 1)$ with $n = 12$ bits.
The linear terms $-2V^Tx$ become biases on the individual bits, rewarding their activation where the data resemble the corresponding template, the diagonal terms of $W$ generate intra-parameter couplings that penalize the simultaneous activation of bits of the same parameter, and the off-diagonal terms generate inter-parameter couplings that penalize activating two components where their templates overlap spatially.

The QUBO is solved with the SQA sampler of OpenJij\footnote{\url{https://www.openjij.org}}, a classical path-integral Monte Carlo emulation of quantum annealing, using 100 reads and 1000 sweeps per run.

\subsection{Localization}
We replace the continuous coordinates by a regular grid of $M$ candidate locations, indexed by $m = 1, ..., M$. Each candidate has a unit-flux source template $S_{ij}^{(m)}$, its own flux amplitude $c_a^{(m)}$ and a binary indicator $z_m \in \{0,1\}$:
\begin{equation}
    \theta_{ij} = g_b B_{ij} + \sum_{m=1}^{M} z_m c_a^{(m)} S_{ij}^{(m)} .
    \label{eq:placeholder_label}
\end{equation}

A single annealing run returns the selected position $\hat{m}$, its flux $\hat{c}_a$, and the background normalization $\hat{g}_b$. To enforce that exactly one candidate is selected, we add a one-hot penalty term $\lambda(\Sigma_m z_m - 1)^2$ to the objective, which penalizes any configuration with more or fewer than one active position. In our runs, we set $\lambda = 5 \cdot max_m TS_m$, i.e. a few times the largest single-candidate TS over the grid, so that the penalty dominates the energy scale of the flux and background terms without overwhelming them. We verified a posteriori that this choice yields exactly one active indicator in all 500 maps, so that no invalid configurations occur.

The QUBO energy provides a direct estimate of the source detection significance. In fact, the comparison of two models reduces to the comparison of two annealing energies.
For each map, we solve two separate QUBO problems, the first one encodes the full model $\theta = g_bB + c_aS^{\hat{m}}$, where its ground-state energy is $E_{best}$. The second one encodes the null hypothesis $\theta_0 = g_bB$ in which only the background is present, its energy is $E_{null}$. The TS is obtained as:
\begin{equation}
    TS = E_{\text{null}} - E_{\text{best}} = \chi^2_{\text{null}} - \chi^2_{\text{best}},
    \label{eq:placeholder_label}
\end{equation}
in analogy with the classical approach. The significance is reported as $\sqrt{TS}$. Since the source position is optimized together with the flux, the TS follows a $\chi^2$ distribution with three degrees of freedom under the null hypothesis \citep{1996ApJ...461..396M}. We note, however, that the three degrees of freedom correspond to the continuous case of \citep{1996ApJ...461..396M}, where the flux and the two coordinates are free parameters. In our formulation the position is instead a discrete choice among 81 candidates, so the $\chi^{2}_3$ distribution should be regarded as an approximation. A precise calibration would require accounting for the fact that the position is selected over a grid of candidates, which we leave to future work.

To identify the energy difference with the likelihood ratio, the annealer must reach the ground state of the problem Hamiltonian, whose energy equals the minimum $\chi^2$. We verified this against the closed-form solution, $E_{null}$ matches the exact minimum, while $E_{best}$ agrees with it to within about $0.5\%$ of the TS over the dataset.

The full localization problem involves $n_{bit}(1 + M) + M$ binary variables, of the order of $10^3$ for our setup, with dense connectivity. The flux bits of each candidate couple to the background bits, and the one-hot indicators are mutually connected. Mapping such a densely connected problem onto the sparse topologies of current quantum annealers (D-Wave's Pegasus and Zephyr) requires minor-embedding, whose physical-qubit cost grows roughly quadratically with the number of logical variables. As a result, an instance of this size would exceed the $\approx 5000$ qubits available on current D-Wave Advantage devices. To run this method on real quantum hardware we will need to reduce the problem size and to use specific tools to find the best embedding. We leave this to future work.

\section{Results}
\label{sec:res}
\begin{figure}[!h]
    \centering
    \includegraphics[width=1\linewidth]{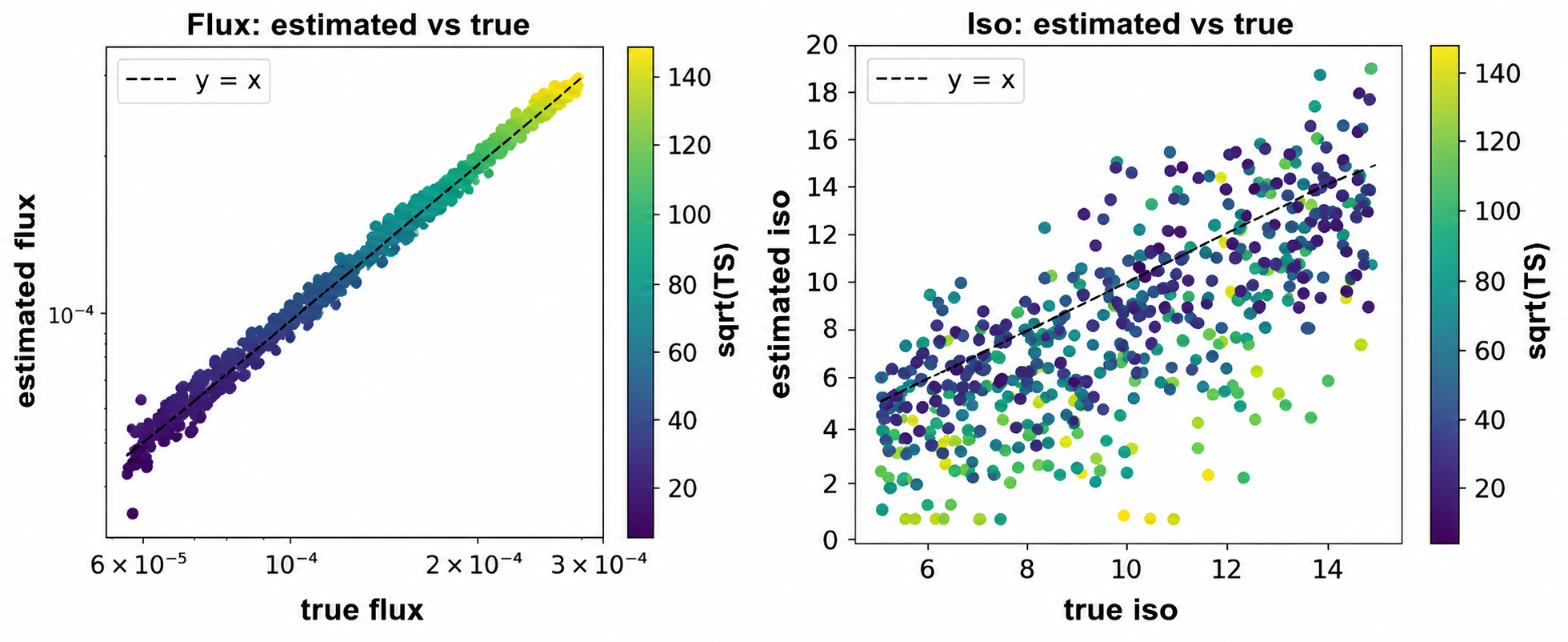}
    \caption{Flux and isotropic estimates from the classical AGILEpy analysis.}
    \label{fig:classical_res}
\end{figure}

Figure \ref{fig:classical_res} shows the AGILEpy estimates against the true values for the source flux (left) and the isotropic normalization (right) \citep{agilepy}. The flux is recovered accurately over the full dynamic range (mean bias of $3.0 \%$, standard deviation of $5.0\%$), while the isotropic term is systematically underestimated (-19.4\%) with a large standard deviation (30.3\%).

\begin{figure}[!h]
    \centering
    \includegraphics[width=1\linewidth]{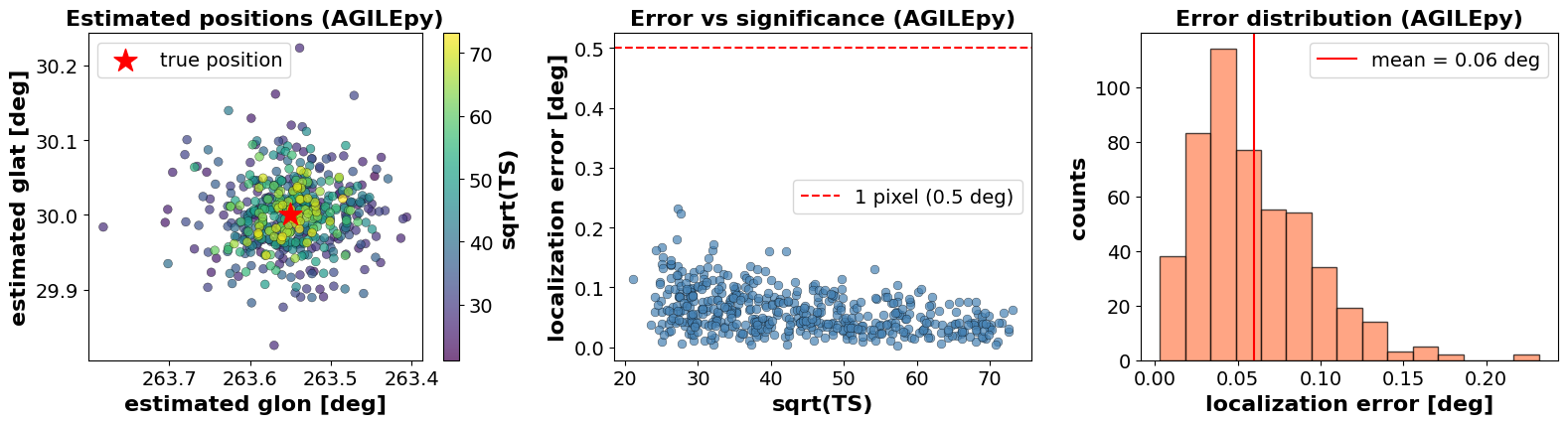}
    \caption{Localization results of the classical AGILEpy analysis.}
    \label{fig:classical_res1}
\end{figure}

\begin{figure}[!h]
    \centering
    \includegraphics[width=1\linewidth]{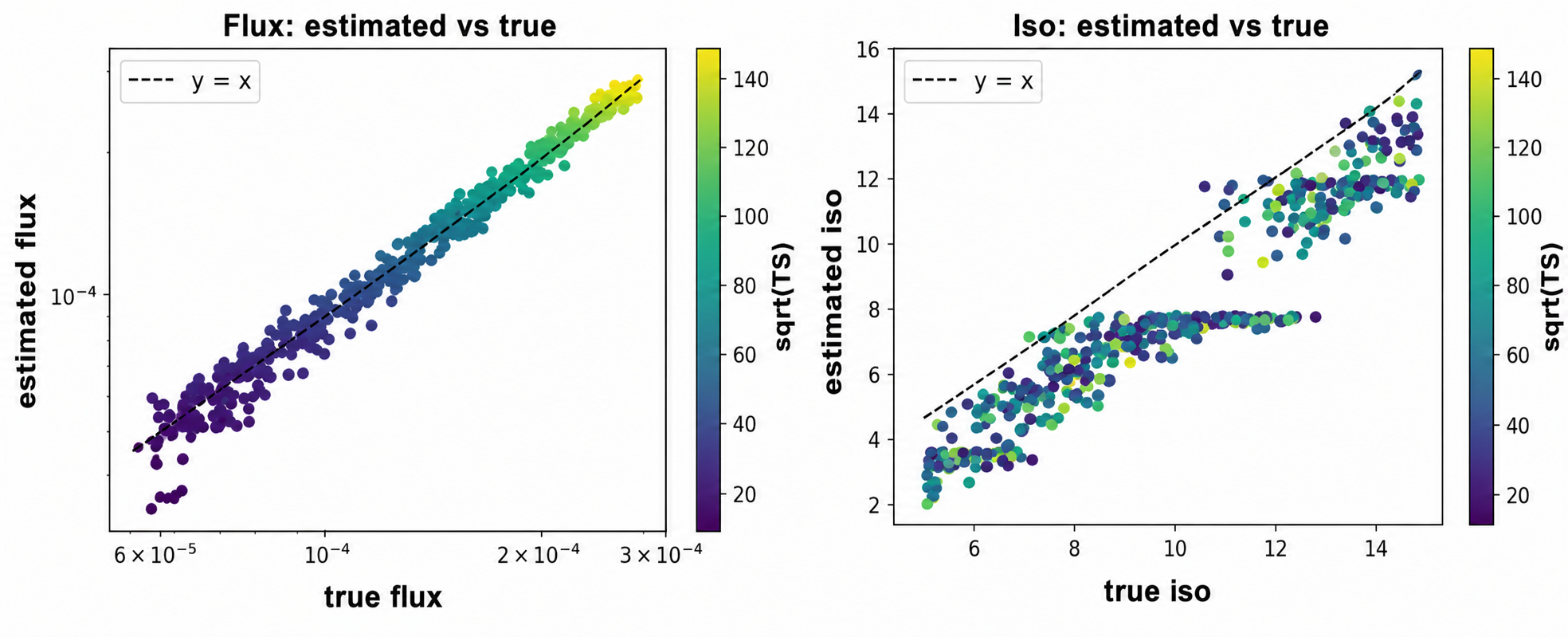}
    \caption{QUBO results against the true values for the source flux (left) and the background (right).}
    \label{fig:annealer}
\end{figure}

The detection significance $\sqrt{TS}$ goes from 21 to 73, with a mean of 43, well above the standard detection threshold, and scales with the source flux as expected in the background-dominated regime. Figure \ref{fig:classical_res1} shows the AGILEpy localization results. The estimated positions cluster around the true source (on the left), while we can notice how the localization error decreases with detection significance. Moreover, all estimates lie within one pixel (0.5°), confirming that the classical fit localizes point sources with optimal accuracy.\\
\newpage
Regarding the quantum annealer, over the 500 simulated maps, the QUBO obtains the source flux with a mean bias of $-1.9\%$ with a standard deviation of $8.3\%$ but underestimates the isotropic normalization by $-20.0\%$, with a standard deviation of $10.8\%$. The isotropic estimate is biased low over the whole range, and its flattening that can be observed in Figure \ref{fig:annealer} reflects the finite range of the fixed-point encoding of the isotropic amplitude.

We found that the QUBO selects the correct node of the candidate grid in 97\% of the maps, the remaining cases being displaced by a single grid step (0.5°). The $\sqrt{TS}$ obtained goes from 17 to 145, with a mean of 65, systematically larger than the classical values. 

\begin{figure}[!h]
    \centering
    \includegraphics[width=0.5\linewidth]{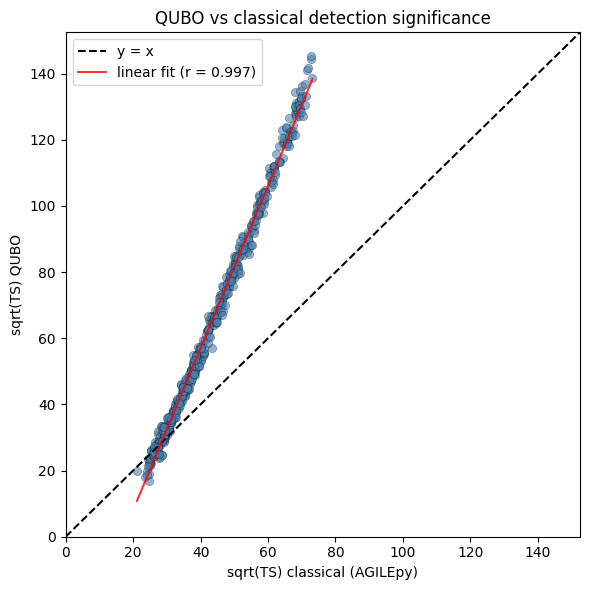}
    \caption{QUBO versus classical detection significance ($\sqrt{TS}$) over the simulated dataset. The two statistics are tightly correlated (r = 0.997). The QUBO values are systematically higher because the unweighted least-squares cost differs from the Poisson likelihood.}
    \label{fig:scatter}
\end{figure}

The QUBO test statistics are systematically larger than the classical ones, but the two are almost perfectly correlated across the dataset (Pearson r =  0.997, Figure \ref{fig:scatter}), following a linear relation over the full range of significances. We verified that this offset does not arise from the spatial region over which the statistic is evaluated. We found that restricting the sum to a region of interest around the source does not change the QUBO TS. In fact, it comes from the cost function itself. The QUBO minimizes an unweighted least-squares $\chi^2$, which assigns unit variance to every pixel, whereas the classical TS derives from the Poisson likelihood. At the low count levels typical of high Galactic latitude, the unweighted $\chi^2$ up-weights low-count pixels, yielding systematically higher TS values.

\section{Conclusions}
\label{sec:conc}
We have shown that the standard maximum-likelihood procedure for gamma-ray source detection can be reformulated as a QUBO problem and solved by simulated quantum annealing. On 500 simulated AGILE-GRID observations, the annealer selected the candidate grid node coinciding with the true source position in 97\% of the maps, and recovered the source flux with a mean bias of -1.9\%, comparable to the classical pipeline. Therefore, these preliminary results indicate that the Gaussian approximation and the binary encoding do not degrade the scientific accuracy of the estimates.

Future work is needed to perform a simultaneous localization of multiple close sources, since the formulation in this work assumes the presence of a single source. Moreover, the systematic offset affecting the isotropic normalization deserves further investigation, as does the behavior of this method on fainter sources, for which the Gaussian approximation is expected to become less accurate. Finally, the QUBO should be tested on actual quantum annealing hardware.
\newpage
\bibliography{sample-ceur}

\end{document}